\documentstyle[11pt,paspconf]{article}

\newcommand{\apg}{\:^{>}_{\sim}\:}
\newcommand{\apl}{\:^{<}_{\sim}\:}
\newcommand{\lya}{\mbox{${\rm Ly}\alpha$}}

\begin{document}

\title{Galaxies of Redshift $z > 5$:  The View from Stony Brook}

\author{Kenneth M. Lanzetta\altaffilmark{1}, Hsiao-Wen Chen\altaffilmark{1},
Alberto Fern\'andez-Soto\altaffilmark{2}, Sebastian Pascarelle\altaffilmark{1},
Noriaki Yahata\altaffilmark{1}, and Amos Yahil\altaffilmark{1}}

\altaffiltext{1}
{Department of Physics and Astronomy, State University of New York at Stony
Brook, Stony Brook, NY 11794-3800, U.S.A.}

\altaffiltext{2}
{Department of Astrophysics and Optics, School of Physics, University of New
South Wales, Kensington--Sydney, NSW 2052, AUSTRALIA}

\begin{abstract}

  We report on some aspects of our efforts to establish properties of the
extremely faint galaxy population by applying our photometric redshift
technique to the HDF and HDF--S WFPC2 and NICMOS fields.  We find that
cosmological surface brightness dimming effects play a dominant role in setting
what is observed at redshifts $z > 2$, that the comoving number density of high
intrinsic surface brightness regions increases monotonically with increasing
redshift, and that previous estimates neglect a significant or dominant
fraction of the ultraviolet luminosity density of the universe due to surface
brightness effects.  The ultraviolet luminosity density of the universe
plausibly increases monotonically with increasing redshift to redshifts beyond
$z = 5$.

\end{abstract}


\section{Introduction}

  We have over the past few years applied our photometric redshift
technique to the Hubble Deep Field (HDF) and Hubble Deep Field South (HDF--S)
WFPC2 and NICMOS fields (e.g.\ Lanzetta, Yahil, \& Fern\'andez-Soto 1996, 1998;
Fern\'andez-Soto, Lanzetta, \& Yahil 1999;  Yahata et al.\ 2000).  Our
objective is to establish properties of the extremely faint galaxy population
by identifying galaxies that are too faint to be spectroscopically identified
by even the largest ground-based telescopes.  Our experiences indicate that
photometric redshift measurements are at least as robust and reliable as
spectroscopic redshift measurements (and probably significantly more so).
Specifically, comparison of photometric and reliable spectroscopic measurements
in the HDF and HDF--S fields demonstrates that the photometric redshift
measurements are accurate to within an RMS relative uncertainty of $\Delta z \
(1+z) \apl 10\%$ and that there are {\em no} known examples of photometric
redshift measurements that are in error by more than a few times the RMS
uncertainty.  These results apply at all redshifts $z < 6$ that have yet been
examined.  It thus appears that the photometric redshift technique provides a
means of obtaining redshift identifications of large samples of the faintest
galaxies to the largest redshifts.

  Here we report on some aspects of our efforts.  Highlights of the results
include the following:

  1.  We have identified nearly 3000 faint galaxies, of which nearly 1000
galaxies are of redshift $z > 2$ and more than 50 galaxies are of redshift $z >
5$ (ranging up to and beyond $z = 10$).  Further, we have fully characterized
the survey area versus depth relationships, in terms of both energy flux
density and surface brightness, in order to measure statistical properties of
the very high redshift galaxy population.

  2.  We find that cosmological surface brightness dimming effects play a 
dominant role in setting what is observed at redshifts $z > 2$.  Most
importantly, we find that it is more or less meaningless to interpret the
galaxy luminosity function (or its moments) at high redshifts without
explicitly taking account of surface brightness effects.

  3.  We find that the comoving number density of high intrinsic surface
brightness regions (or in other words of high star formation rate density
regions) increases monotonically with increasing redshift.

  4.  We find that previous estimates neglect a significant of dominant
fraction of the ultraviolet luminosity density of the universe due to surface
brightness effects and that the rest-frame ultraviolet luminosity density (or
equivalently the cosmic star formation rate density) has not yet been measured
at redshifts $z \apg 2$.  The ultraviolet luminosity density of the universe
plausibly increases monotonically with increasing redshift to redshifts beyond
$z = 5$.

  The most recent versions of our photometry and redshift catalogs of faint
galaxies in the HDF and HDF--S fields can be found on our web site at:
\begin{center}
{\tt http://www.ess.sunysb.edu/astro/hdfs/}.
\end{center}
Here and throughout we adopt a standard Friedmann cosmological model of
dimensionless Hubble constant $h = H_0 / (100 \ {\rm km} \ {\rm s}^{-1} \ {\rm
Mpc}^{-1})$ and deceleration parameter $q_0 = 0.5$.

\section{Observations and Analysis}

  Our current observations and analysis differ from our previous observations
and analysis in three important ways:

  First, we have included all available public ground- and space-based imaging
observations of the HDF, HDF--S WFPC2, and HDF--S NICMOS fields.  Details of
the current observations are summarized in Table 1.

\begin{center}
\begin{tabular}{p{1.75in}l}
\multicolumn{2}{c}{Table 1} \\
\hline
\hline
\multicolumn{1}{c}{Field} & \multicolumn{1}{c}{Filters} \\
\hline
HDF \dotfill          & F300W, F450W, F606W, F814W, \\
& F110W, F160W, $J$, $H$, $K$ \\
HDF--S WFPC2 \dotfill  & F300W, F450W, F606W, F814W, \\
& $U$, $B$, $V$, $R$, $I$, $J$, $H$, $K$ \\
HDF--S NICMOS \dotfill & F110W, F160W, F222M, STIS, \\
& $U$, $B$, $V$, $R$, $I$ \\
\hline
\end{tabular}
\end{center}

  Second, we have developed and applied a new quasi-optimal photometry
technique based on fitting models of the spatial profiles of the objects (which
are obtained using a non-negative least squares image reconstruction method) to
the ground- and space-based images, according to the spatial profile fitting
technique described previously by Fern\'andez-Soto, Lanzetta, \& Yahil (1999).
For faint objects, the signal-to-noise ratios obtained by this technique are
larger than the signal-to-noise ratios obtained by aperture photometry
techniques by typically a factor of two.

  Third, we have measured photometric redshifts using a sequence of six
spectrophotometric templates, including the four templates of our previous
analysis (of E/S0, Sbc, Scd, and Irr galaxies) and two new templates (of
star-forming galaxies).  Inclusion of the two new templates eliminates the
tendency of our previous analysis to systematically underestimate the redshifts
of galaxies of redshift $2 < z < 3$ (by a redshift offset of roughly 0.3), in
agreement with results found previously by Ben\'{\i}tez et al.\ (1999).

  The accuracy and reliability of the photometric redshift technique is
illustrated in Figure 1, which shows the comparison of 108 photometric and
reliable spectroscopic redshifts in HDF and HDF--S.  (Note that a
non-negligible fraction of published spectroscopic redshift measurements of
galaxies in HDF and HDF--S have been shown to be in error and so must be
excluded from consideration.)  With the sequence of six spectrophotometric
templates, the photometric redshifts are accurate to within an RMS relative
uncertainty of $\Delta z/(1 + z) \apl 10\%$ and there are {\em no} known
examples of photometric redshift measurements that are in error by more than a
few times the RMS uncertainty.  These results apply at all redshifts $z < 6$
that have yet been examined.  Details of some of our current observations and
analysis are described by Yahata et al.\ (2000).

\begin{figure}[t]
\includegraphics{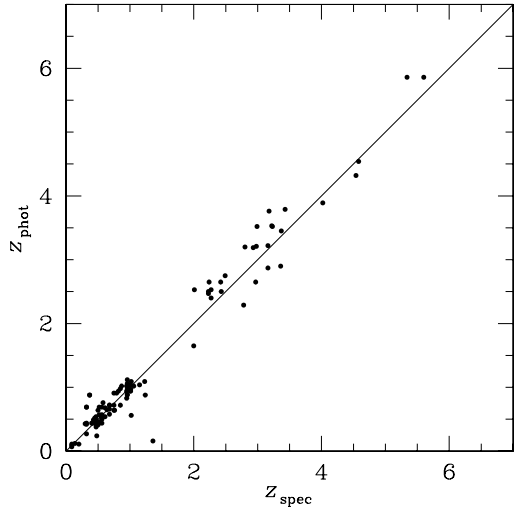}
\vspace{2.65in}
\caption{Comparison of 108 photometric and reliable spectroscopic measurements
of galaxies in HDF and HDF--S.  The RMS dispersion between the photometric and
reliable spectroscopic measurements is $\approx 0.1$ at $z < 2$, $\approx 0.3$
at $2 < z < 4$, and $\approx 0.15$ at $z > 4$.}
\end{figure}

\section{Stony Brook Faint Galaxy Redshift Survey}

  Our analysis of the HDF and HDF--S WFPC2 and NICMOS fields constitutes a
survey of galaxies to the faintest energy flux density and surface brightness
limits currently accessible.  Properties of the redshift survey are as follows:

  First, we have determined nine- or 12-band photometric redshifts of faint
galaxies in three deep fields.

  Second, we have selected galaxies at both optical and infrared wavelengths,
in two or more of the F814W, F160W, $H$, and $K$ bands (depending on field).
(We have related selection in different bands by adopting the spectral energy
distribution of a star-forming galaxy).

  Third, we have fully characterized the survey area versus depth relations, as
functions of both energy flux density and surface brightness.

  Fourth, we have established properties of the extremely faint galaxy
population by using a maximum-likelihood parameter estimation technique and a
bootstrap resampling parameter uncertainty estimation technique.  The derived
parameter uncertainties explicitly account for the effects of photometric
error, sampling error, and cosmic dispersion with respect to the
spectrophotometric templates.

  The Stony Brook faint galaxy redshift survey includes nearly 3000 faint
galaxies, of which nearly 1000 galaxies are of redshift $z > 2$ and more than
50 galaxies are of redshift $z > 5$ (ranging up to and beyond $z = 10$).  The
depth and scope of the survey is summarized in Figure 2, which shows redshift
distributions of all galaxies identified in the HDF and HDF--S WFPC2 and NICMOS
fields.  The redshift distributions of galaxies identified in the HDF and
HDF--S WFPC2 field are characterized by broad peaks at redshift $z \approx 1$
and long tails extending to redshifts $z > 5$.  Further, the distributions are
statistically different from one another (with the HDF--S WFPC2 field
exhibiting a statistically significant excess of galaxies of redshift $z > 2$
compared with the HDF), and both exhibit statistically significant large-scale
fluctuations.  The redshift distribution of galaxies identified in the HDF--S
NICMOS field is characterized by a broad peak at redshift $z \approx 1$ and a
long tail extending to redshifts $z > 10$.

\begin{figure}[t]
\includegraphics{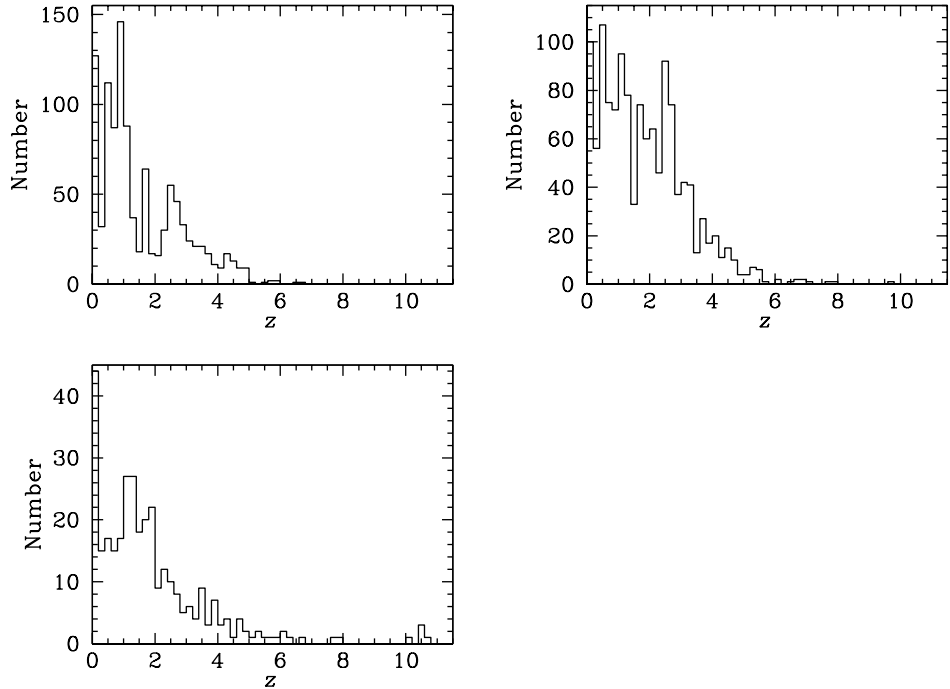}
\vspace{3.75in}
\caption{Redshift distributions of all galaxies identified in the HDF (upper
left) and HDF--S WFPC2 (upper right) and NICMOS (lower left) fields.}
\end{figure}

\section{Some High-Redshift Galaxies}

  Examples of some high-redshift galaxies are shown in Figure 3, which plots
observed and modeled spectral energy distributions and redshift likelihood
functions of galaxies identified in the HDF--S WFPC2 and NICMOS fields.

\begin{figure}
\includegraphics{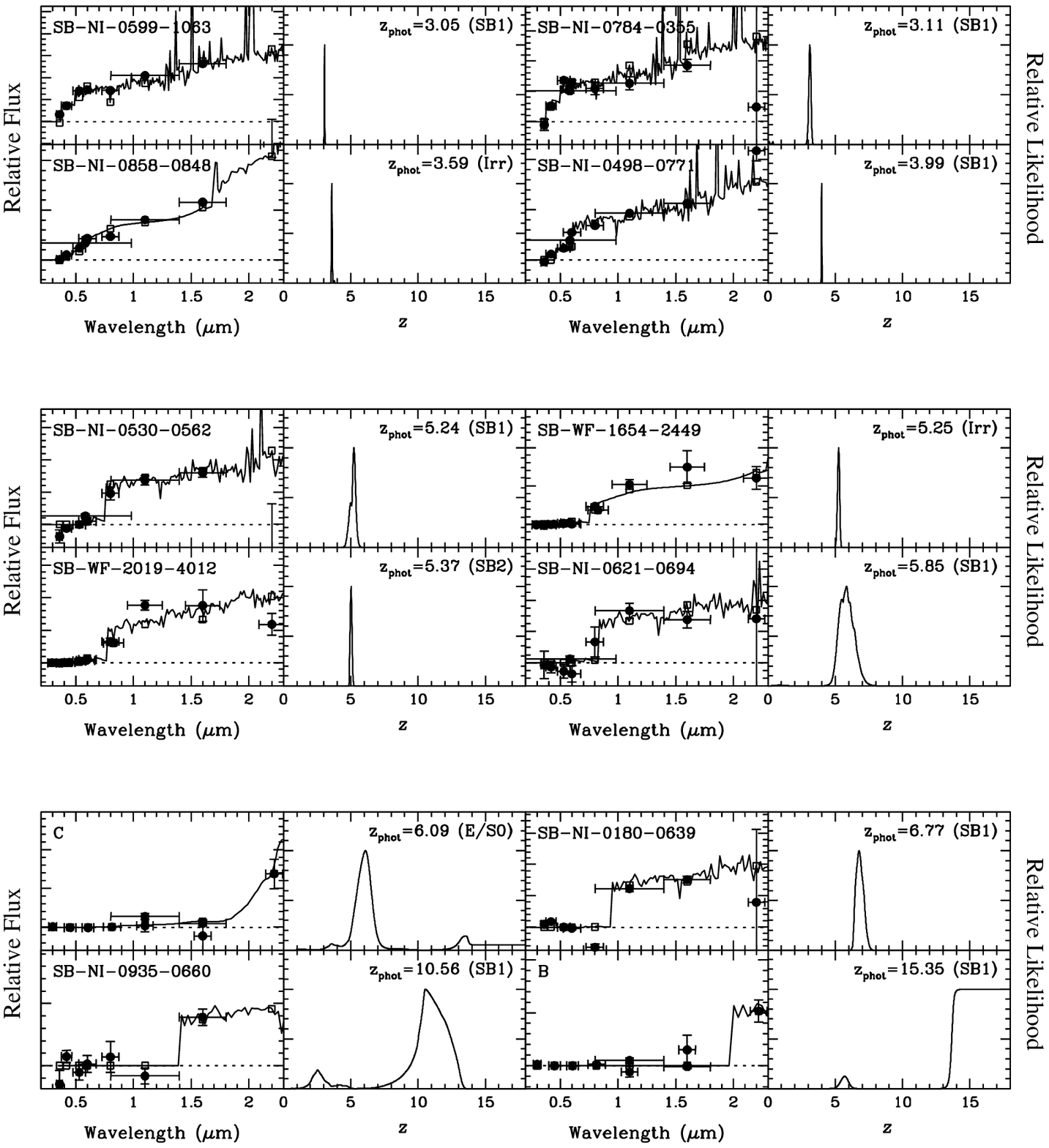}
\vspace{6.25in}
\caption{Observed and modeled spectral energy distributions (left-most of each
pair of panels) and redshift likelihood functions (right-most of each pair of
panels) of galaxies identified in the HDF--S.  Top group of panels shows
galaxies of redshift $3 < z < 4$, middle group of panels shows galaxies of
redshift $5 < z < 6$, and bottom group of panels shows galaxies of redshift
$z > 6$.  Filled circles are measured fluxes and open squares are best-fit
model fluxes.  Vertical error bars indicate $1 \sigma$ uncertainties and
horizontal error bars indicate filter FWHM.}
\end{figure}

  The top group of panels of Figure 3 shows four galaxies of redshift $3 < z <
4$ and near-infrared continuum magnitude $AB(8140) \approx 25$, and the middle
group of panels of Figure 3 shows four galaxies of redshift $5 < z < 6$ and
near-infrared continuum magnitude $AB(8140) \approx 26$.  In each case, the
spectral enery distribution shows unambiguous evidence of the \lya-forest and
Lyman-limit decrements, and the redshift likelihood function is very sharply
peaked, indicating that, of all the spectrophotometric models considered, the
appropriately redshifted spectrophotometric template provides the only
plausible fit to the observations.  We believe that the redshifts indicated in
the top and middle groups of panels of Figure 3 are established with
essentially complete certainty---and with substantially greater certainty than
has been or could be achieved by means of spectroscopic observations of
galaxies of the same redshifts and continuum magnitudes.  The galaxies shown in
the top and middle groups of panels of Figure 3 are unexceptional, and results
shown for these galaxies are completely representative of results obtained for
other similar galaxies.

  Results of Figure 1 indicate that at redshifts $3 < z < 4$, the RMS
measurement uncertainty of the photometric redshift technique is $\Delta z
\approx 0.3$ or $\Delta z / (1 + z) \approx 10\%$, which we believe results
primarily due to stochastic variations in the density of the \lya\ forest among
different lines of sight.  Results of Figure 1 indicate (albeit with limited
statistical certainty) that at redshifts $5 < z < 6$ the RMS measurement
uncertainty of the photometric redshift technique is $\Delta z \approx 0.15$ or
$\Delta z / (1 + z) \approx 3\%$, which we believe is superior to results at
redshifts $3 < z < 4$ because almost complete absorption in the \lya\ forest
allows for less stochastic variations in the density of the \lya\ forest among
different lines of sight.

  The bottom group of panels of Figure 3 shows four galaxies of best-fit
photometric redshift measurement $z > 6$ and near-infrared continuum magnitude
$AB(16,000) \approx 27$, including two galaxies (galaxies B and C) that we
identified previously as candidate extremely high redshift galaxies on the
basis of ground-based near-infrared measurements (Lanzetta, Yahil, \&
Fern\'andez-Soto 1998).  At these redshifts and continuum magnitudes, the
redshift determinations are not unambiguous, and the high-redshift solutions
are typically accompanied by lower-redshift solutions, of early-type galaxies
of redshift $z \approx 3$.  Additional deep imaging observations of these
galaxies are needed to establish their redshifts with certainty.

\section{The Galaxy Luminosity Function at Redshifts $z > 2$}

  We have modeled the rest-frame 1500 \AA\ luminosity function of galaxies of
redshift $z > 2$ by adopting an evolving Schechter luminosity function
\begin{equation}
\Phi(L,z) = \Phi_* / L_*(z) [ L / L_*(z) ]^{-\alpha} \exp[ -L / L_* (z)]
\end{equation}
with
\begin{equation}
L_*(z) = L_*(z = 3) \left( \frac{1 + z}{4} \right)^\beta.
\end{equation}
The best-fit parameters for a simultaneous fit to the HDF and HDF--S WFPC2 and
NICMOS fields (where we have related selection in different bands by adopting
the spectral energy distribution of a star-forming galaxy) are $\Phi_* = 0.004
\pm 0.001 \ h^3$ Mpc$^{-3}$, $L_* = 2.7 \pm 0.3 \times 10^{28} \ h^{-2}$ erg
s$^{-1}$ Hz$^{-1}$, $\alpha = 1.49 \pm 0.03$, and $\beta = -1.2 \pm 0.3$.  The
best-fit model is compared with the observations in Figure 4, which shows the
cumulative galaxy surface density versus redshift and magnitude for galaxies
selected in the F814W and F160W bands.

  From a practical point of view, Figure 4 presents our best measurements and
models of the empirical galaxy surface density versus redshift and
near-infrared magnitude.  The most striking result of Figure 4 is that galaxies
identified by our analysis at the highest redshifts $z > 7$ (which are detected
only at the faintest F160W magnitudes $AB \apg 28$) are predicted by a
straightforward extrapolation of a plausible model of the high-redshift galaxy
luminosity function.  For our analysis to have uncovered {\em no} galaxies of
redshift $z > 7$ would have implied rapid evolution of the galaxy luminosity
function at redshifts $z > 6$.
    
\begin{figure}[t]
\includegraphics{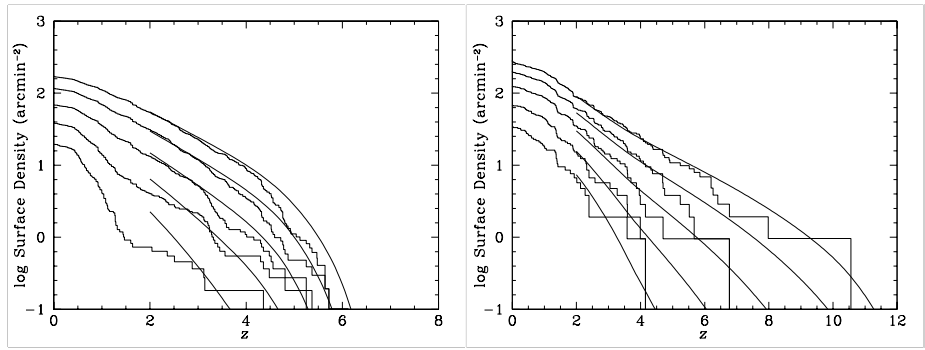}
\vspace{1.65in}
\caption{Logarithm of cumulative galaxy surface density versus redshift and
magnitude (i.e.\ surface density of galaxies of redshift greater than a given
redshift) for galaxies selected in the F814W (left panel) and F160W (right
panel) bands.  Smooth curves are best-fit model, and jagged curves are
observations.  Different curves show different magnitude thresholds, ranging
from $AB = 24$ (bottom curves) through $AB = 28$ (top curves).}
\end{figure}

\section{Effects of Cosmological Surface Brightness Dimming}

  Results of the previous section indicate that the galaxy luminosity function
evolves only mildly at redshifts $z > 2$, i.e.\ as $(1 + z)^\beta$ with $\beta
\approx -1$.  But due to $(1 + z)^3$ cosmological surface brightness dimming,
the measured luminosities of extended objects decrease rapidly with increasing
redshift, even if the actual luminosities of the objects remain constant.  For
this reason, we consider it more or less meaningless to interpret the galaxy
luminosity function (or its moments) over a redshift interval spanning $z = 2$
through $z = 10$ without explicitly taking account of surface brightness
effects.

  To make explicit the effects of cosmological surface brightness dimming on
observations of high-redshift galaxies, we have constructed the ``star
formation rate intensity distribution function'' $h(x)$.  Specifically, we
consider all pixels contained within galaxies on an individual pixel-by-pixel
basis.  Given the redshift of a pixel (which is set by the photometric redshift
of the host galaxy), an empirical $k$ correction (which is set by the model
spectral energy distribution of the host galaxy) and a cosmological model
determine the rest-frame 1500 \AA\ luminosity of the pixel, and an angular
plate scale and a cosmological model determine the proper area of the pixel.
Adopting a Salpeter initial mass function to convert the rest-frame 1500 \AA\
luminosity to the star formation rate and dividing the star formation rate by
the proper area yields the ``star formation rate intensity'' $x$ of the pixel.
Summing the proper areas of all pixels within given star formation rate
intensity and redshift intervals, dividing by the star formation rate intensity
interval, and dividing by the comoving volume then yields the ``star formation
rate intensity distribution function,'' which we designate as $h(x)$.  The star
formation rate intensity distribution function $h(x)$ is exactly analogous to
the QSO absorption line systems column density distribution function $f(N)$ (as
a function of neutral hydrogen column density $N$).  In terms of the star
formation rate intensity distribution function, the unobscured cosmic star
formation rate density $\dot{\rho}_s$ (or equivalently the rest-frame
ultraviolet luminosity density) is given by
\begin{equation}
\dot{\rho}_s = \int_0^\infty x h(x) dx.
\end{equation}

  Results are shown in Figure 5, which plots the star formation rate intensity
distribution function $h(x)$ versus star formation rate intensity $x$
determined from galaxies identified in the HDF and HDF--S NICMOS field.
Several results are apparent on the basis of Figure 5:  First, the star
formation rate intensity threshold of the survey is an extremely strong
function of redshift, ranging from $x_{\rm min} \approx 5 \times 10^{-4}$
$M_\odot$ yr$^{-1}$ kpc$^{-2}$ at $z \approx 0.5$ to $x_{\rm min} \approx 1$
$M_\odot$ yr$^{-1}$ kpc$^{-2}$ at $z \approx 6$.  {\em We conclude that
cosmological surface brightness dimming effects play a dominant role in setting
what is observed at redshifts $z > 2$.}  Second, the comoving number density of
high intrinsic surface brightness regions increases monotonically with
increasing redshift.  {\em We conclude that the comoving number density of 
high intrinsic surface brightness regions (or equivalently of high star
formation rate density regions) increases monotonically with increasing
redshift}.  (See also Pascarelle, Lanzetta, \& Fern\'andez-Soto 1998).  Third,
at redshifts $z \apl 1.5$ [at which $h(x)$ is measured over a wide range in
$x$], the distribution is characterized by a relatively shallow slope at $\log
x \apl -1.5$ $M_\odot$ yr$^{-1}$ kpc$^{-2}$ and by a relatively steep slope at
$\log x \apg -1.5$ $M_\odot$ yr$^{-1}$ kpc$^{-2}$.  These slopes are such that
the bulk of the cosmic star formation rate density occurs at $\log x \approx
-1.5$ $M_\odot$ yr$^{-1}$ kpc$^{-2}$, which is measured only at redshifts $z
\apl 2$.  {\em We conclude that previous estimates neglect a significant or
dominant fraction of the ultraviolet luminosity density of the universe due to
surface brightness effects and that the rest-frame ultraviolet luminosity
density (or equivalently the cosmic star formation rate density) has not yet
been measured at redshifts $z \apg 2$.}

\begin{figure}[t]
\includegraphics{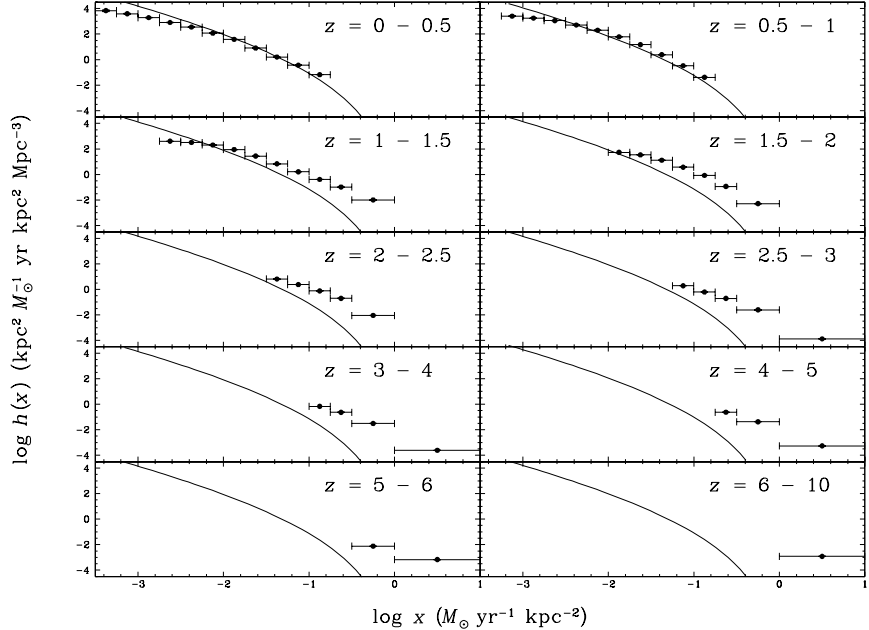}
\vspace{3.40in}
\caption{Logarithm of star formation rate intensity distribution function
$h(x)$ versus logarithm of star formation rate intensity $x$, determined from
galaxies identified in the HDF and HDF--S NICMOS field.  Different panels show
different redshift intervals, ranging from $z = 0$ through 10.  Points show
observations, with vertical error bars indicating $1 \sigma$ uncertainties and
horizontal error bars indicating bin sizes.  Smooth curves show a fiducial
model (based on a bulge spatial profile) adjusted to roughly match the
observations at $z = 0 - 0.5$.}
\end{figure}

  This last point is illustrated in Figure 6, which shows the ultraviolet
luminosity density of the universe versus redshift measured to various
intrinsic surface brightness thresholds.  Specifically, Figure 6 shows the
ultraviolet luminosity density of the universe versus redshift measured to
intrinsic surface brightness thresholds that could be detected in the HDF at
all redshifts to $z = 5.0$, to $z = 3.4$, to $z = 2.3$, to $z = 1.6$, and to $z
= 1.1$.  (Higher intrinsic surface brightness thresholds can be seen to higher
redshifts, whereas lower intrinsic surface brightness thresholds can be seen
only to lower redshifts.)  Results of Figure 6 indicate that to any fixed
intrinsic surface brightness threshold, the ultraviolet luminosity density of
the universe increases monotonically with increasing redshift.  Apparently, the
ultraviolet luminosity density of the universe plausibly increases
monotonically with increasing redshift to redshifts beyond $z = 5$.

\begin{figure}[t]
\includegraphics{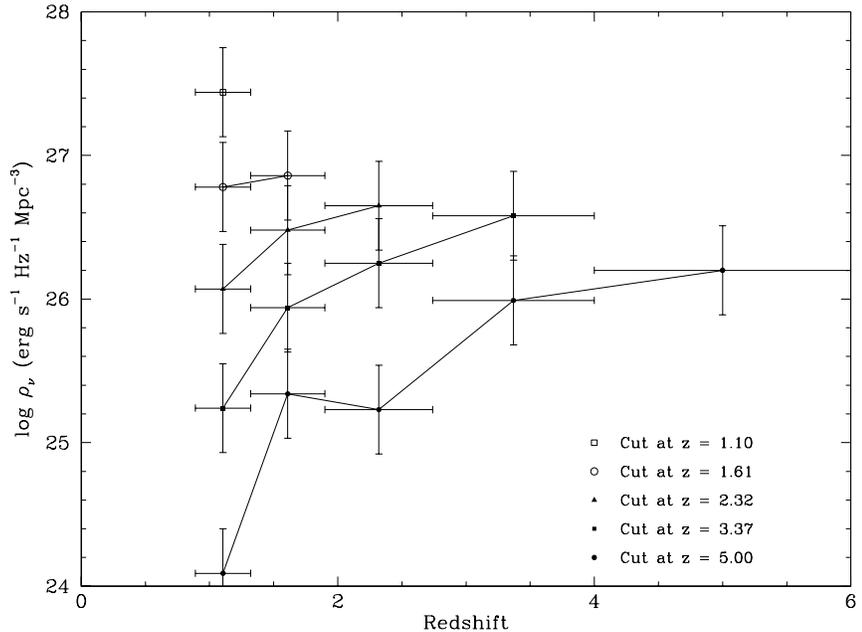}
\vspace{3.50in}
\caption{Logarithm of ultraviolet luminosity density versus redshift measured
to various intrinsic surface brightness thresholds, determined from galaxies
identified in the HDF.}
\end{figure}

\acknowledgments

  We thank Hy Spinrad and Daniel Stern for providing spectroscopic redshift
measurements in advance of publication and acknowledge Mark Dickinson and Roger
Thompson for obtaining NICMOS observations of HDF.  This research was supported
by NASA grant NACW--4422 and NSF grant AST--9624216 and is based on
observations with the NASA/ESA Hubble Space Telescope and on observations
collected at the European Southern Observatory.


\begin{references}

\reference Ben\`{\i}tez, N, Broadhurst, T., Bouwens, R., Silk, J., \& Rosati,
P., 1999, ApJ, 515, L65

\reference Fern\'andez-Soto, A., Lanzetta, \& Yahil, A. 1999, ApJ, 513, 34

\reference Lanzetta, K. M., Yahil, A., \& Fern\'{a}ndez-Soto, A. 1996, Nature,
381, 759

\reference \underline{\makebox[0.5in]{}}. 1998, AJ, 116, 1066

\reference Pascarelle, S., Lanzetta, K. M., \& Fern\'andez-Soto, A. 1998, ApJ,
508, L1

\reference Yahata, N., Lanzetta, K. M., Chen, H.-W., Fern\'andez-Soto, A.,
Pascarelle, S., Puetter, R., \& Yahil, A. 2000, ApJ, submitted

\end{references}
\end{document}